\documentclass[twocolumn,showpacs,prb]{revtex4}
\usepackage{graphicx}
\usepackage{dcolumn}
\usepackage{amsmath}
\usepackage{amssymb}
\usepackage{times}

\makeatletter
\def\btt#1{\texttt{\@backslashchar#1}}
\DeclareRobustCommand\bblash{\btt{\@backslashchar}} \makeatother

\begin{document}

\title[Short Title]{Josephson Effect due to Odd-frequency Pairs in Diffusive
Half Metals}
\author{Yasuhiro Asano$^1$, Yukio Tanaka$^{2,3}$ and Alexander A. Golubov$^4$%
}
\affiliation{$^1$Department of Applied Physics, Hokkaido University, Sapporo 060-8628,
Japan\\
$^2$Department of Applied Physics, Nagoya University, Nagoya 464-8603, Japan%
\\
$^3$CREST, Japan Science and Technology Corporation (JST) Nagoya, 464-8603,
Japan \\
$^4$Faculty of Science and Technology, University of Twente, 7500 AE,
Enschede, The Netherlands }
\date{\today}

\begin{abstract}
The Josephson effect in superconductor / diffusive ferromagnet /
superconductor (SFS) junctions is studied using the recursive Green
function method in the regime of large exchange energy in a
ferromagnet. Motivated by recent experiment [R.~S.~Keizer,
\textit{et. al.,} Nature \textbf{439}, 825 (2006)] we also address
the case of superconductor / diffusive half metal / superconductor
junctions. The pairing function in spin-singlet and triplet
channels, the Josephson current and their mesoscopic fluctuations
are calculated. We show that the spin-flip scattering at the
junction interfaces opens the Josephson channel of the odd-frequency
spin-triplet Cooper pairs. As a consequence, the local density of
states in half metals has a large peak at the Fermi energy.
Therefore odd-frequency pairs can be detected experimentally
by using the scanning tunneling spectroscopy.
\end{abstract}

\maketitle

Ferromagnetism and spin-singlet superconductivity are competing orders
against each other because the exchange field breaks down the spin-singlet
pairs. The Cooper pairs, however, do not always disappear under exchange
fields. The Fulde-Ferrell-Larkin-Ovchinnikov~\cite{fulde,larkin} (FFLO)
state and the proximity effect in ferromagnets~\cite%
{buzdin,buzdin2,petrashov,ryazanov,kontos,golubov,bergeret,kadigrobov} are
typical examples. Under exchange fields, the pairing function oscillates and
changes its sign in the real space. As a consequence, superconductor /
ferromagnet / superconductor (SFS) junctions undergo the 0-$\pi $ transition
with varying length of a ferromagnet or temperature. It was also predicted
that the odd-frequency spin-triplet pairs appear in weakly polarized
ferromagnets with rotating magnetization direction near the junction
interface~\cite{bergeret}. These pairs have long range penetrarion
into ferromagnets~\cite{bergeret,kadigrobov}.
Thus Cooper pairs change their
original face to survive under exchange fields.

Half metal is an extreme case of completely spin polarized material because
its electronic structure is insulating for one spin direction and metallic
for the other. At a simple thought, the spin-singlet Cooper pairs would not
be able to penetrate into half metals. However recent experiment~\cite%
{keizer} showed the existence of the Josephson coupling in superconductor /
half metal / superconductor (S/HM/S) junctions, where NbTiN was used as a $s$%
-wave superconducting electrode and CrO$_{2}$ as a half metal. Thus one has
to seek a new state of Cooper pairs in half metals attached to spin-singlet
superconductors. Prior to the experiment~\cite{keizer}, Eschrig\textit{\ et.
al.~}\cite{eschrig} have addressed this challenging issue. In the
\emph{clean limit}, they showed that the $p$-wave spin-triplet pairs induced by
the spin-flip scattering at the interface can carry the Josephson current.
In real S/HM/S junctions, however, half metals are in the diffusive
transport regime; the elastic mean free path of an electron is much smaller
than the size of a half metal. In addition, real S/HM/S junctions are close
to the dirty limit because the coherence length in a ferromagnet may become
comparable to the mean free path. In such diffusive half metals, the
$p$-wave symmetry of Cooper pairs is not possible because the pair wave
function is isotropic in the momentum space due to impurity
scattering~\cite{tanaka06}.
In this paper, we provide a general theory for the Josephson effect in
diffusive SFS junctions with arbitrary magnitude of the exchange field $V_{ex}$.
When $V_{ex}$ is much larger than the superconducting pair
potential at zero temperature $\Delta_0$,
mesoscopic fluctuations of the Josephson
current are much larger than the ensemble averaged value.
In addition, we focus on interesting case of diffusive S/HM/S junctions.
We show that the odd-frequency spin-triplet $s$-wave pairing state is realized
in half metals and propose an experimental method to detect this property.

\begin{figure}[tbh]
\begin{center}
\includegraphics[width=8.0cm]{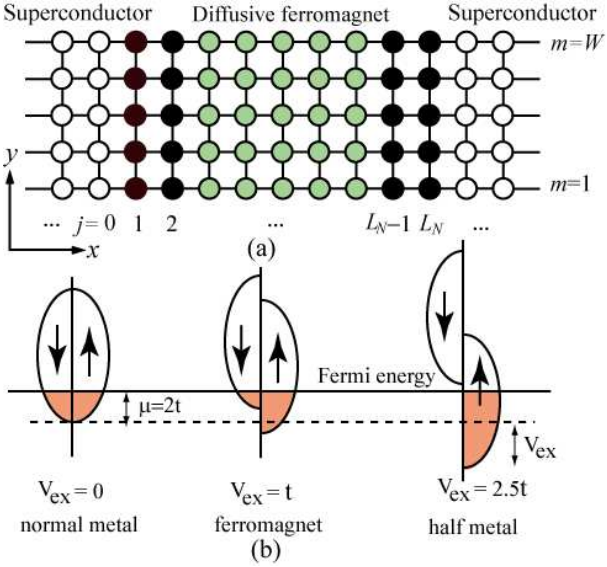}
\end{center}
\caption{ (Color online) (a): A schematic figure of a SFS junction on the
tight-binding lattice. (b): The density of states for each spin direction.
The Josephson junction is of the SNS, SFS, and S/HM/S type for $V_{ex}/t=0$,
1 and 2.5, respectively. }
\label{fig1}
\end{figure}

Let us consider the two-dimensional tight-binding model for a SFS junction
as shown in Fig.~\ref{fig1}(a). The vector $\boldsymbol{r}=j{\boldsymbol{x}}%
+m{\boldsymbol{y}}$ points to a lattice site, where ${\boldsymbol{x}}$ and ${%
\boldsymbol{y}}$ are unit vectors in the $x$ and $y$ directions,
respectively. In the $y$ direction, we apply the periodic boundary condition
for the number of lattice sites being $W$.
Electronic states in superconducting junctions are described by the
mean-field Hamiltonian
\begin{align}
H_{\text{BCS}}=& \frac{1}{2}\sum_{\boldsymbol{r},\boldsymbol{r}^{\prime }}%
\left[ \tilde{c}_{\boldsymbol{r}}^{\dagger }\;\hat{h}_{\boldsymbol{r},%
\boldsymbol{r}^{\prime }}\;\tilde{c}_{\boldsymbol{r}^{\prime }}^{{}}-%
\overline{\tilde{c}_{\boldsymbol{r}}}\;\hat{h}_{\boldsymbol{r},\boldsymbol{r}%
^{\prime }}^{\ast }\;\overline{\tilde{c}_{\boldsymbol{r}^{\prime }}^{\dagger
}}\;\right]   \notag \\
& +\frac{1}{2}\sum_{\boldsymbol{r}\in \text{S}}\left[ \tilde{c}_{%
\boldsymbol{r}}^{\dagger }\;\hat{\Delta}\;\overline{\tilde{c}_{\boldsymbol{r}%
}^{\dagger }}-\overline{\tilde{c}_{\boldsymbol{r}}}\;\hat{\Delta}^{\ast }\;%
\tilde{c}_{\boldsymbol{r}}\right] ,  \label{bcs}
\end{align}%
\begin{align}
\hat{h}_{\boldsymbol{r},\boldsymbol{r}^{\prime }}=& \left[ -t\delta _{|%
\boldsymbol{r}-\boldsymbol{r}^{\prime }|,1}+(\epsilon _{\boldsymbol{r}}-\mu
+4t)\delta _{\boldsymbol{r},\boldsymbol{r}^{\prime }}\right] \hat{\sigma}_{0}
\notag \\
-& \boldsymbol{V}(\boldsymbol{r})\cdot \hat{\boldsymbol{\sigma}}\delta _{%
\boldsymbol{r},\boldsymbol{r}^{\prime }},
\end{align}%
with $\overline{\tilde{c}}_{\boldsymbol{r}}=\left( c_{\boldsymbol{r}%
,\uparrow },c_{\boldsymbol{r},\downarrow }\right) $, where $c_{\boldsymbol{r}%
,\sigma }^{\dagger }$ ($c_{\boldsymbol{r},\sigma }^{{}}$) is the creation
(annihilation) operator of an electron at $\boldsymbol{r}$ with spin $\sigma
=$ ( $\uparrow $ or $\downarrow $ ), $\overline{\tilde{c}}$ means the
transpose of $\tilde{c}$, $\hat{\sigma}_{l}$ for $l=1-3$ are the Pauli's
matrices, and $\hat{\sigma}_{0}$ is $2\times 2$ unit matrix. The hopping
integral $t$ is considered among nearest neighbor sites in both
superconductors and ferromagnets. In a ferromagnet, the on-site scattering
potentials are given randomly in the range of $-V_{I}/2\leq \epsilon _{%
\boldsymbol{r}}\leq V_{I}/2$ and the uniform exchange potential is given by $%
\boldsymbol{V}(\boldsymbol{r})=V_{ex}\boldsymbol{e}_{3}$, where $%
\boldsymbol{e}_{l}$ for $l=1-3$ is unit vector in a spin space. The Fermi
energy $\mu $ is set to be $2t$ in a normal metal with $V_{ex}=0$, while a
ferromagnet and a half metal are respectively described by $V_{ex}/t$ = 1
and 2.5 in Fig.~\ref{fig1}(b). The spin-flip scatterings are introduced at $%
j=1,2$, $L_{N}-1$, and $L_{N}$, where we choose $\boldsymbol{V}(%
\boldsymbol{r})=V_{S}\boldsymbol{e}_{2}$. In superconductors we take $%
\epsilon _{\boldsymbol{r}}=0$ and choose $\hat{\Delta}=i\Delta \hat{\sigma}%
_{2}$, where $\Delta $ is the amplitude of the pair potential in the $s$%
-wave symmetry channel.

The Hamiltonian is diagonalized by the Bogoliubov transformation and the
Bogoliubov-de Gennes (BdG) equation is numerically solved by the recursive
Green function method~\cite{furusaki,ya01-1}. We calculate the Matsubara
Green function,
\begin{equation}
\check{G}_{\omega _{n}}(\boldsymbol{r},\boldsymbol{r}^{\prime })=\left(
\begin{array}{cc}
\hat{g}_{\omega _{n}}(\boldsymbol{r},\boldsymbol{r}^{\prime }) & \hat{f}%
_{\omega _{n}}(\boldsymbol{r},\boldsymbol{r}^{\prime }) \\
-\hat{f}_{\omega _{n}}^{\ast }(\boldsymbol{r},\boldsymbol{r}^{\prime }) & -%
\hat{g}_{\omega _{n}}^{\ast }(\boldsymbol{r},\boldsymbol{r}^{\prime })%
\end{array}
\right) , \label{deff}
\end{equation}
where $\omega _{n}=(2n+1)\pi T$ is the Matsubara frequency, $n$ is an
integer number, and $T$ is a temperature. The Josephson current is given by
\begin{equation}
J=-ietT\sum_{\omega _{n}}\sum_{m=1}^{W}\mathrm{Tr}\left[ \check{G}_{\omega
_{n}}(\boldsymbol{r}^{\prime },\boldsymbol{r})-\check{G}_{\omega _{n}}(%
\boldsymbol{r},\boldsymbol{r}^{\prime })\right]
\end{equation}
with $\boldsymbol{r}^{\prime }=\boldsymbol{r}+\boldsymbol{x}$. In this
paper, $2\times 2$ and $4\times 4$ matrices are indicated by $\hat{\cdots}$
and $\check{\cdots}$, respectively. The quasiclassical Green function method
is a powerful tool to study the proximity effect. However the quasiclassical
Green function cannot be constructed in a half metal because the Fermi
energy is no longer much larger than the pair potential for one spin
direction. On the other hand, there is no such difficulty in our method. In
addition, it is possible to obtain the ensemble average of the Josephson
current $\langle J\rangle =(1/N_{S})\sum_{i=1}^{N_{S}}J_{i}$ and its
fluctuations $\delta J=\sqrt{\langle J^{2}\rangle -\langle J\rangle ^{2}}$
after calculating Josephson current in a large number of samples $N_{S}$
with different impurity configurations. These are the advantages of the
recursive Green function method. Throughout this paper we fix the following
parameters: $L_{N}=74$, $W=25$, $\mu =2t$, $V_{I}=2t$ and $\Delta _{0}=0.005t
$~[\onlinecite{length}]. This parameter choice corresponds to the diffusive
transport regime in the N, F and HM layers. The results
presented below are not sensitive to variations of these
parameters.

We first discuss the Josephson current in SFS junctions as shown in Fig.~\ref%
{fig2}(a) for $T=0.1T_{c}$ where $T_{c}$ is the transition temperature.
We assume that the
spin-flip scattering at the interfaces is absent (i.e., $V_{S}=0$) and fix the
phase difference across the junctions $\varphi $ equal to $\pi /2$.
\begin{figure}[tbh]
\begin{center}
\includegraphics[width=9.5cm]{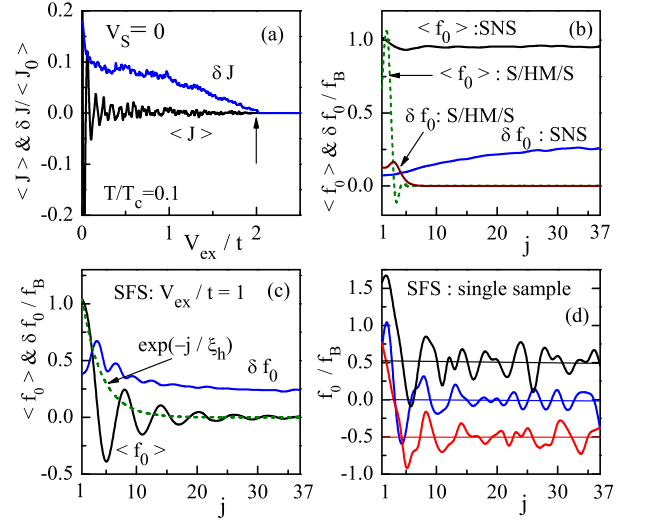}
\end{center}
\caption{ (Color online) (a) The Josephson current versus the exchange
potential $V_{ex}$. At $V_{ex}=2t$ ferromagnets become half-metallic as
indicated by an arrow.  (b) The pairing function of spin-singlet pairs and
its fluctuations versus position $j$ for $V_{ex}/t=0$ (SNS) and $2.5$ (S/HM/S);
(c) for $V_{ex}/t=1$ (SFS).
(d) The
pairing function in three different samples for $V_{ex}/t=1$, the vertical
axis is offset by 0.5 as indicated by the horizontal lines. The spin-flip
scattering is absent in all panels, $V_{S}=0$. }
\label{fig2}
\end{figure}
The results are normalized by $\langle J_{0}\rangle $ which is the Josephson
current in superconductor / normal metal / superconductor (SNS) junctions
(i.e., $V_{ex}=0$).
We define the coherence length $\xi_h=\sqrt{D/2V_{ex}}/a_0$ measured in units
of lattice constant $a_0$ with $D$ being the diffusion coefficient.
The Josephson current oscillates as a function of $V_{ex}$
and changes its sign almost periodically. The sign changes of $\langle
J\rangle $ correspond to the 0-$\pi $ transitions in an SFS junction. At the
same time, the amplitude of $\langle J\rangle $ decreases rapidly with
increasing $V_{ex}$. We should pay attention to the relation $\langle
J\rangle \ll \delta J$ which means that the Josephson current is not the
self-averaging quantity. It is impossible to predict the Josephson current
in a single sample $J_{i}$ from the ensemble average $\langle J\rangle $
because $J_{i}$ strongly depends on a microscopic impurity configuration. In
fact, the Josephson current flows in \emph{a single sample} even if $\langle
J\rangle =0$ at the transition points. Roughly speaking, $\langle J\rangle $
vanishes because half of samples are 0-junctions and the rest are $\pi $%
-junctions~\cite{ya01-2}. Since $\langle J\rangle =0$, $\delta J$
approximately corresponds to the typical amplitude of the Josephson current
expected in a single sample. The relation $\langle J\rangle =0$ has
different meaning for SFS and S/HM/S cases. In SFS junctions, $\langle
J\rangle =0$ at the transition points is the result of the ensemble
averaging and the Josephson current remains finite in a single sample. The
characteristic temperature and length of a ferromagnet at the $0-\pi $
transitions vary from one sample to another. In S/HM/S junctions at
$V_{ex}=2.5t$, however, $\langle J\rangle =0$ means vanishing Josephson
current even in a single sample~\cite{eschrig} because
$\langle J\rangle=\delta J=0$.

The origin of large fluctuations of the Josephson current can be understood
by considering the behavior of the pairing function in a ferromagnet. The
pairing function in Eq.~(\ref{deff}) can be decomposed into four components,
\begin{equation}
\frac{1}{W}\sum_{m=1}^{W}\hat{f}_{\omega _{n}}(\boldsymbol{r},\boldsymbol{r}%
)=i\sum_{\nu =0}^{3}f_{\nu }(j)\hat{\sigma}_{\nu }\hat{\sigma}_{2},
\label{dec}
\end{equation}
where $f_{0} (f_3)$ is the pairing function of the spin-singlet (spin-triplet) pairs
with the spin structure of $(\left\vert \uparrow \downarrow \right\rangle
-(+)\left\vert \downarrow \uparrow \right\rangle )/\sqrt{2}$, respectively,
and the pairing
function of $\left\vert \upuparrows \right\rangle $ $(\left\vert
\downdownarrows \right\rangle )$ is given by $f_{\uparrow \uparrow
}=if_{2}-f_{1}$ ($f_{\downarrow \downarrow }=if_{2}+f_{1}$). In Figs.~\ref%
{fig2}(b) and (c), we show $\langle f_{0}\rangle $ and $\delta f_{0}$ as a
function of position $j$ in a diffusive ferromagnet, where $f_{B}$ is the
pairing function in bulk superconductor, $\omega _{n}$ is fixed at $%
0.02\Delta _{0}$, $V_{S}=0$, and $\varphi =0$. The junction interface and
the center of a ferromagnet correspond to $j=1$ and $j=37$, respectively. In
SNS junctions (Fig.~\ref{fig2}b), $\langle f_{0}\rangle $ is larger than $%
\delta f_{0}$ and very weakly decays with $j$. The spin-singlet Cooper pairs
exist everywhere in a normal metal. Near the interface, $\delta f_{0}$ is
slightly suppressed due to the tight contact to the superconductor. On the
other hand, in SFS junctions in (Fig.~\ref{fig2}c), the average $\langle
f_{0}\rangle $ decreases exponentially with $j$ according to
$\exp (-j/\xi_{h})$ as indicated by a broken line.
The fact that $\delta f_{0}$ remains finite at the center of a
ferromagnet means that the spin-singlet pairs penetrate
far beyond $\xi _{h}$ even though
$\langle f_{0}\rangle \sim 0$ there.

In Fig.~\ref{fig2}(d), we show the pairing function for three different
realization of disorder in SFS junctions. The pairing functions are
in phase near the interface ($j\leq \xi _{h}$), whereas they are out of
phase far from the interface. We obtain the relation
$\delta f_{0}\propto e^{-j/\xi _{T}}$ with $\xi _{T}=\sqrt{D/2\omega_{n}}$
in agreement with Ref.~\onlinecite{zyuzin}. Thus we conclude that spin-singlet
Cooper pairs do exist in \emph{a single sample} of ferromagnet even for $%
j\gg \xi _{h}$ and the mesoscopic fluctuations of the pairing function
provide the origin of the large fluctuations in the Josephson current. In
S/HM/S junctions for $V_{ex}=2.5t$ as shown in Fig.~\ref{fig2}(b), both $%
\langle f_{0}\rangle $ and $\delta f_{0}$ vanish for $j\gg 1$, which
indicates the absence of spin-singlet Cooper pairs in a half metal.

\begin{figure}[tbh]
\begin{center}
\includegraphics[width=9.5cm]{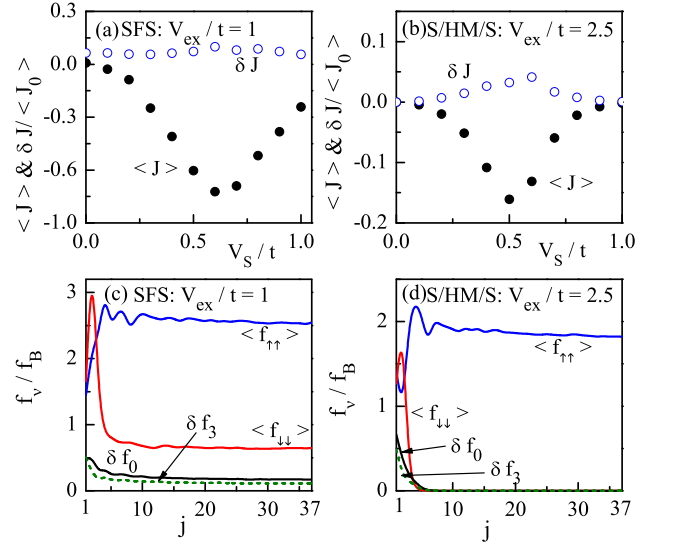}
\end{center}
\caption{ (Color online) (a) The Josephson current and its fluctuations at $%
T=0.1T_{c}$ and $\protect\varphi =\protect\pi /2$ as a function of the
interface spin-flip scattering $V_{S}$ for $V_{ex}/t=1$ and (b) for $%
V_{ex}/t=2.5$. The pairing functions versus position $j$ in a ferromagnet (c) and in a half
metal (d) at $V_{s}=0.4t,$ $\protect\varphi =0$ and $\protect\omega %
_{n}=0.02\Delta _{0}$. In (d), differences among $\langle f_{\downarrow
\downarrow }\rangle $, $\protect\delta f_{0}$ and $\protect\delta f_{3}$
at large $j$ become small. }
\label{fig3}
\end{figure}
The relation $\langle J\rangle \ll \delta J$ is the characteristic feature
of the Josephson current in diffusive SFS junctions. This feature, however,
is drastically changed by the spin-flip scattering at the interfaces. In
Figs.~\ref{fig3} (a) and (b) we show $\langle J\rangle $ and $\delta J$
\textit{vs} $V_{S}$ for $V_{ex}/t=1$ and 2.5, respectively. In both cases
(a) and (b), we find that $|\langle J\rangle |\geq \delta J$ for $V_{S}\geq
0.3t$. The Josephson current becomes self-averaging in the presence of the
spin-flip scattering. The reason can be explained by calculating the pairing
functions of equal-spin pairs shown in Figs.~\ref{fig3}(c) and (d), where $%
f_{\nu }$ is plotted as a function of position $j$. Here we show
$\delta f_{0}$ and $\delta f_{3}$ instead of $\langle f_{0}\rangle $
and $\langle f_{3}\rangle $ because the ensemble averages are much
smaller than their fluctuations.
The fast decay of $\delta f_0$ and $\delta f_3$ is determined by
strong spin polarization. In both cases (c) and (d), $\langle
f_{\uparrow \uparrow }\rangle $ becomes larger than $\delta f_{0}$ and $%
\delta f_{3}$ because the pairing function $
f_{\uparrow \uparrow }$ does not change sign for various impurity configurations.
Thus the averaged quantities become larger
than their fluctuations. Thus the Josephson current becomes self-averaging
as shown in Figs.~\ref{fig3}(a) and (b).
\begin{figure}[tbh]
\begin{center}
\includegraphics[width=9.5cm]{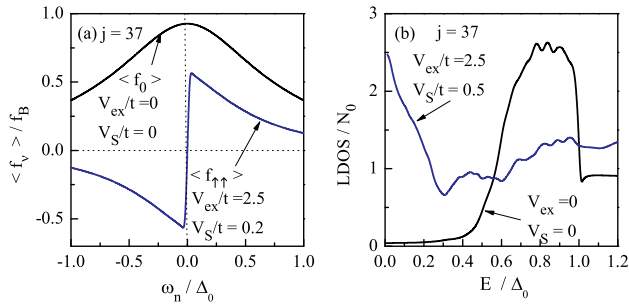}
\end{center}
\caption{ (Color online) (a): Dependences of the pairing functions on
$\protect\omega _{n}$. (b): The local density of states at $j=37$
in a half metal at $V_{S}=0.5t$ and in a normal metal
at $V_{S}=0$. }
\label{fig4}
\end{figure}

Finally we address an unusual symmetry property of the Josephson current in S/HM/S
junctions. In Fig.~\ref{fig4} (a), we show $\langle f_{\uparrow \uparrow
}\rangle $ as a function of $\omega _{n}$, where $j=37$, $V_{S}=0.2t$,
$\varphi =0$, and $V_{ex}=2.5t$. For comparison, we also show $\langle
f_{0}\rangle $ on the normal side of a SNS junction. The pairing function $%
\langle f_{0}\rangle $ in a normal metal is the even function of $\omega _{n}
$, whereas $\langle f_{\uparrow \uparrow }\rangle $ in a half metal is the
odd function of $\omega _{n}$~Ref.~\onlinecite{bergeret}.
The pairing function obeys the Pauli's rule
\begin{equation}
\hat{f}_{\omega _{n}}(\boldsymbol{r},\boldsymbol{r}^{\prime })=-\bar{\hat{f}}%
_{-\omega _{n}}(\boldsymbol{r}^{\prime },\boldsymbol{r}),
\end{equation}%
where $\bar{\hat{f}}$ denotes the transpose of $\hat{f}$ meaning the
interchange of spins. It is well known that ordinary even-frequency pairs
are classified into two symmetry classes: the spin-singlet even-parity and
the spin-triplet odd-parity one. In the former case, the negative sign
arises due to the interchange of spins, while in the latter case due to $%
\boldsymbol{r}\leftrightarrow \boldsymbol{r}^{\prime }$ . In the present
calculation, all components on the right hand side of Eq.~(\ref{dec}) have
the $s$-wave symmetry. The pairing functions are isotropic in both the real
and momentum spaces due to diffusive impurity scattering. As a result, $%
f_{\uparrow \uparrow }$ must be the odd function of $\omega _{n}$ to obey
the Pauli's rule. Both even- and odd-frequency pairs are mixed in
ferromagnets as shown in Fig.~\ref{fig3}(c). The fraction of odd-frequency
pairs depends on parameters such as the exchange potential and the spin-flip
scattering. On the other hand, in a diffusive half metal all Cooper pairs
have the odd-frequency character, which causes drastic change in the
quasiparticle density of states.

The density of states is given by $N(E,j)=-\frac{1}{\pi }\frac{1}{W}%
\sum_{m=1}^{W}\text{Im}\text{Tr}\check{G}_{E+i\gamma }(\boldsymbol{r},%
\boldsymbol{r})$, where $\gamma $ is a small imaginary part chosen to be 0.05%
$\Delta _{0}$ in the following. In Fig.~\ref{fig4}(b), the local density of
states (LDOS) at $j=37$ is shown, where $\varphi =0$ and $N_{0}$ is the
density of states in the normal state at $V_{ex}=0$. For comparison, we show
LDOS on the normal side of SNS junction with $V_{S}=0$ \ which has a minigap
at $E<E_{Th}\sim 0.3\Delta _{0}$, where $E_{Th}$ is the Thouless energy. In
contrast to that, LDOS in a half metal has a peak at the Fermi energy, it's
width is characterized by $E_{Th}$.
This peak is generated at the spin active interface by the mechanism
discussed in Ref~\onlinecite{fogelstrom} and is tranfered into a half metal
due to long range property of odd-frequency spin-triplet even-parity pairing
function.
The peak is much stronger than the enhancement of the
LDOS found in weak ferromagnets~\cite{buzdin,golubov,kontos2,yokoyama}.
In addition, in a half metal
the peak shape is almost independent of position, while in the SF junctions%
\cite{kontos2} the LDOS has an oscillatory peak/dip structure at $E=0$ which
rapidly decays with the distance from the SF interface. Therefore the large
peak at $E=0$ in LDOS is a robust and direct evidence of the odd-frequency
pairing in half metals. To test the existence of such peculiar pairing state,
the scanning
tunneling spectroscopy could be used.


In conclusion, we have studied Josephson effect in superconductor /
diffusive ferromagnet / superconductor junctions by using the
recursive Green function method. The Josephson current in these junctions
basically is not self-averaging because the spin-singlet Cooper pairs
penetrating into ferromagnets far beyond $\xi _{h}$ cause the large
fluctuations of the pairing function.
In the presence of the spin-flip scattering at the interfaces,
the equal-spin odd-frequency pairs drastically suppress the fluctuations.
When ferromagnets are half-metallic, all Cooper pairs have the odd-frequency
property.
As a result, the low energy peak in the quasiparticle density of states
in a half metal exists at the distances far beyond $\xi_h$ from
the interface and could be probed by
scanning tunneling spectroscopy.

We acknowledge helpful discussions with J.Aarts, T.M.Klapwijk,
Yu.Nazarov and A.F.Volkov. This work was partially supported by the
Dutch FOM, the NanoNed program under grant TCS7029 and Grant-in-Aid for Scientific Research on Priority Area "Physics of new quantum phases in superclean materials"
(Grant No. 18043001) from The Ministry of Education, Culture, Sports, 
Science and Technology of Japan.



\begin{thebibliography}{99}
\bibitem{length} The mean free path $\ell$ and the coherence length $\xi_0$
are about 6 and 100 lattice constant, respectively. Qualitatively the same
results as Figs.~2-4 are confirmed also for $\xi_0\sim 10$.
\bibitem{f3} In the presence of the exchange potential, function Im$(f_{3})$
appears in addition to Re$(f_{0})$. The characteristic behavior of Im$(f_{3})
$ is similar to that of Re$(f_{0})$ because $f_{3}$ is also the pairing
function of the hetero spin pairs.
\end{thebibliography}
\end{document}